\documentclass[a4paper, conference]{IEEEtran}
\IEEEoverridecommandlockouts
\usepackage{cite}
\usepackage{amsmath,amssymb,amsfonts}
\usepackage{algorithmic}
\usepackage{graphicx}
\usepackage{textcomp}
\usepackage{xcolor}
\usepackage{textcomp}
\def\BibTeX{{\rm B\kern-.05em{\sc i\kern-.025em b}\kern-.08em
  T\kern-.1667em\lower.7ex\hbox{E}\kern-.125emX}}
\begin{document}

\title{Ge coated silicon nanowires as human respiratory sensing device}

\author{
\IEEEauthorblockN{
\textbf{ E. Fakhri}\IEEEauthorrefmark{1}
\textbf{, M. T. Sultan}\IEEEauthorrefmark{2}
\textbf{, A. Manolescu}\IEEEauthorrefmark{1}
\textbf{, S. Ingvarsson}\IEEEauthorrefmark{2}
\textbf{, H. G. Svavarsson}\IEEEauthorrefmark{1}
}
\IEEEauthorblockA{
{\IEEEauthorrefmark{1}Department of Engineering, Reykjavik University, Menntavegur 1, IS-102 Reykjavik, Iceland}\\
{\IEEEauthorrefmark{2}Science Institute, University of Iceland, Reykjavík, Iceland}
}
}

\maketitle

\begin{abstract}
We report on Ge coated silicon nanowires (SiNWs) sensors synthesized with metal assisted chemical etching and qualify their functionality as human respiratory sensor. The sensors were made from p-type single-crystalline (100) silicon wafers using a silver catalysed top-down etching, afterwards coated by 50 nm Ge thin layer using a magnetron sputtering. The Ge post-treatment were performed by rapid thermal annealing (RTA) at 450 and 700$^{\circ}$C. The sensors were characterized by X-ray diffraction diffractogram and scanning electron microscopy. It is demonstrated that the sensors are highly sensitive as human breath detectors,  with rapid response and frequency detect-ability. They are also shown to be a good candidate for human respiratory diseases diagnoses.
\end{abstract}


\section{Introduction}
Sleep apnea syndrome is a potentially serious disorder during which breathing readily stop and starts with the occurrence of approximately 30 apneas during 7- 8 hour sleep. Increasing number of individuals suffer from sleep apnea complications for instance hypertension and stroke, making sleep apnea the second most occurring cause of stroke in patient \cite{lou2020human}. Therefore, respiratory monitoring of individuals posing sleep apnea syndrome is of prime importance. 

In this context pressure and humidity sensors are of particular interest as they offer broad application in industry process control, environmental monitoring, cleanrooms, medial/ health care facilities and more \cite{Wang2021, ghosh2021fabrication, Liu2018}. With rapid growth in technology, a vast variety of materials have been investigated for pressure and humidity sensing, to mention silicon nanowires (SiNWs), ZnO, GaN,TiO$_2$, carbon nanotubes, composite fibres/polymers(such as PDMS) combined with conductive nanostructure such as metals, carbon nanotubes and modified graphene\cite{Wang2021, ghosh2021fabrication, lou2020human}. However for these composite systems the long them sensing is not reliable as they tend to degrade in ambient atmosphere over time. Of these materials SiNWs pose superior characteristics due to its superior unique structural, electrical, optical, and thermoelectric properties \cite{fakhri2021synthesis,heris2020thermoelectric}, device compatibility, having larger surface area and pose high sensitivity to pressure and humidity.  

It has been shown that SiNWs exhibit anomolous piezoresistance effect munch higher than bulk silicon \cite{lugstein2010anomalous} and several sensors based on piezoresistance properties of SiNWs, prepared with metal assisted chemical etching (MACE) have also been proposed, for instance flexible pressure sensors \cite{Kim2020}, airflow sensing devices \cite{zhang2012piezoresistive}, and nanorod breath sensor \cite{ghosh2021fabrication}. Zhang et al \cite{zhang2013characterizations}, reported a flow sensor based on SiNWs NEMS sensors for a low pressure range which is critical for bio-compatible sensors such as breath sensors. Additionally, stability, fast response and recovery time and stabilized resistance baseline are important factors that most of the proposed sensors could not achieve \cite{lou2020human}.
In this paper we report a low-cost, easy to fabricate, bio-compatible and rapidly-scalable Ge-deposited SiNWs based sensor for their application in human respiratory sensing.

\begin{figure}{b}
    \centering
    \includegraphics[width=0.65\linewidth]{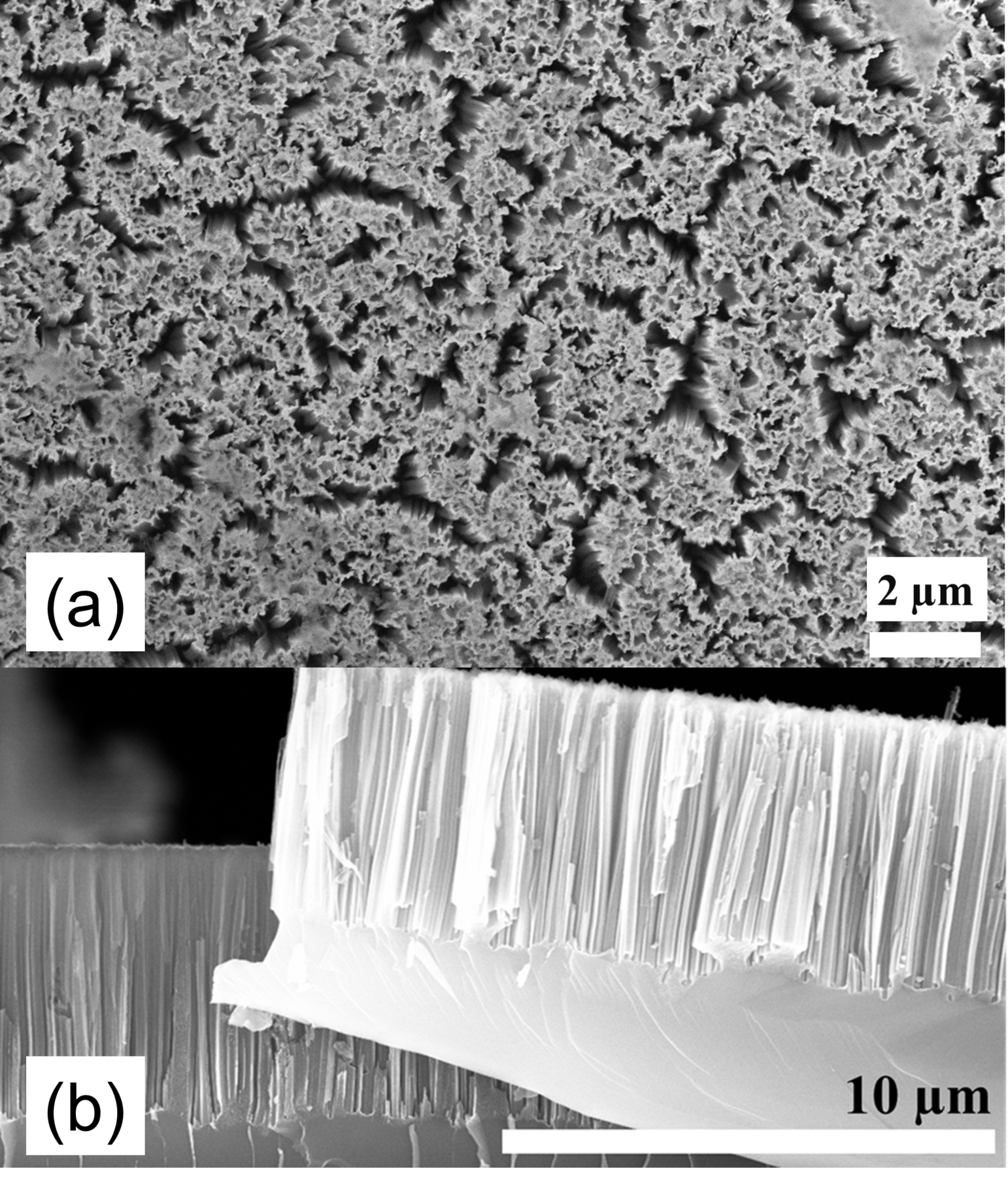}
    \caption{(a) Top-view and (b) cross-sectional SEM micrograph of SiNWs obtained by MACE.}
    \label{1}
\end{figure}

\begin{figure}
    \centering
    \includegraphics[width=0.9\linewidth]{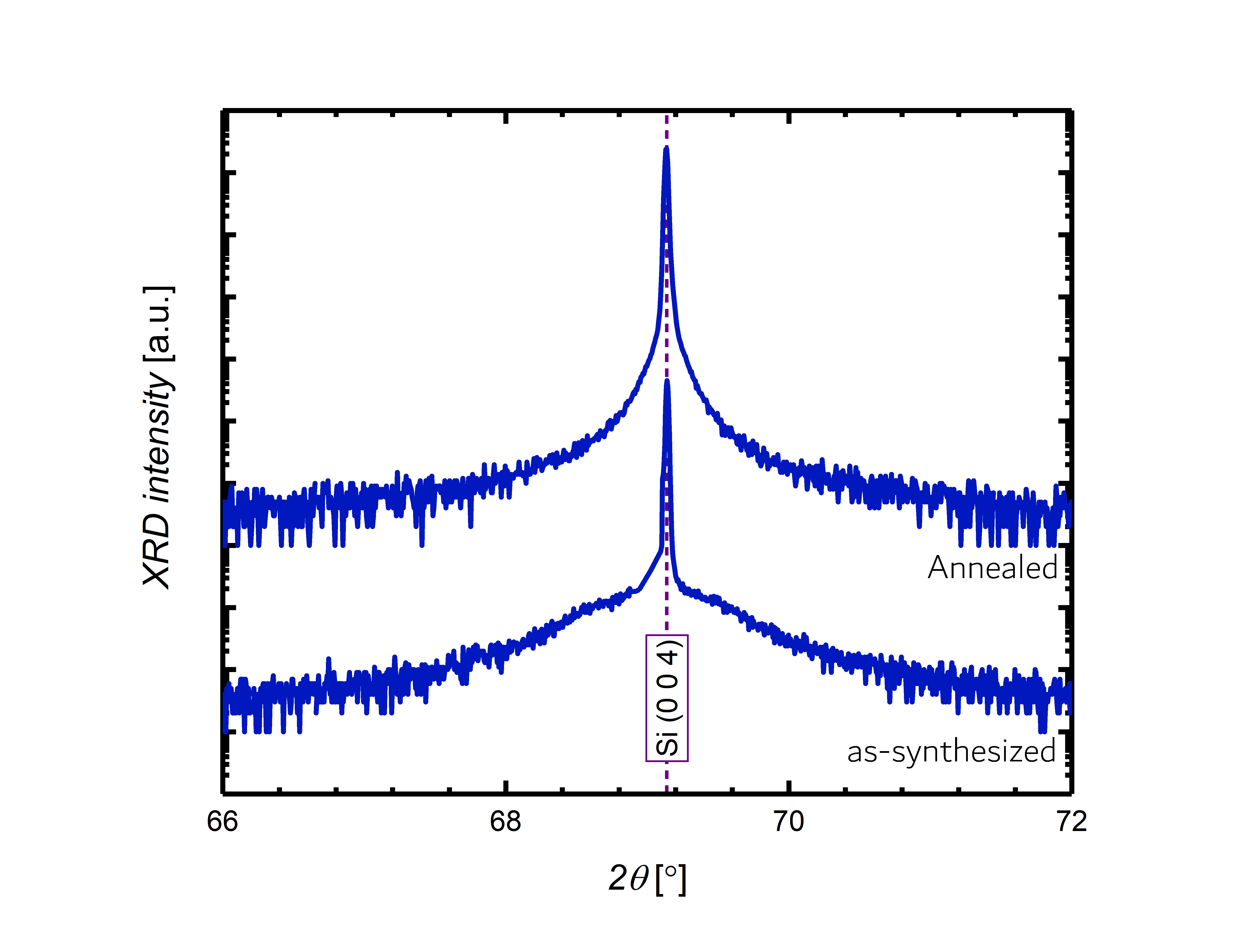}
    \caption{XRD diffractorgram along Si (004) atomic plane obtained over SiNWs based structures both in un-annealed (lower plot) and annealed (upper plot) state.}
    \label{2}
\end{figure}
\begin{figure}
    \centering
    \includegraphics[width=0.75\linewidth]{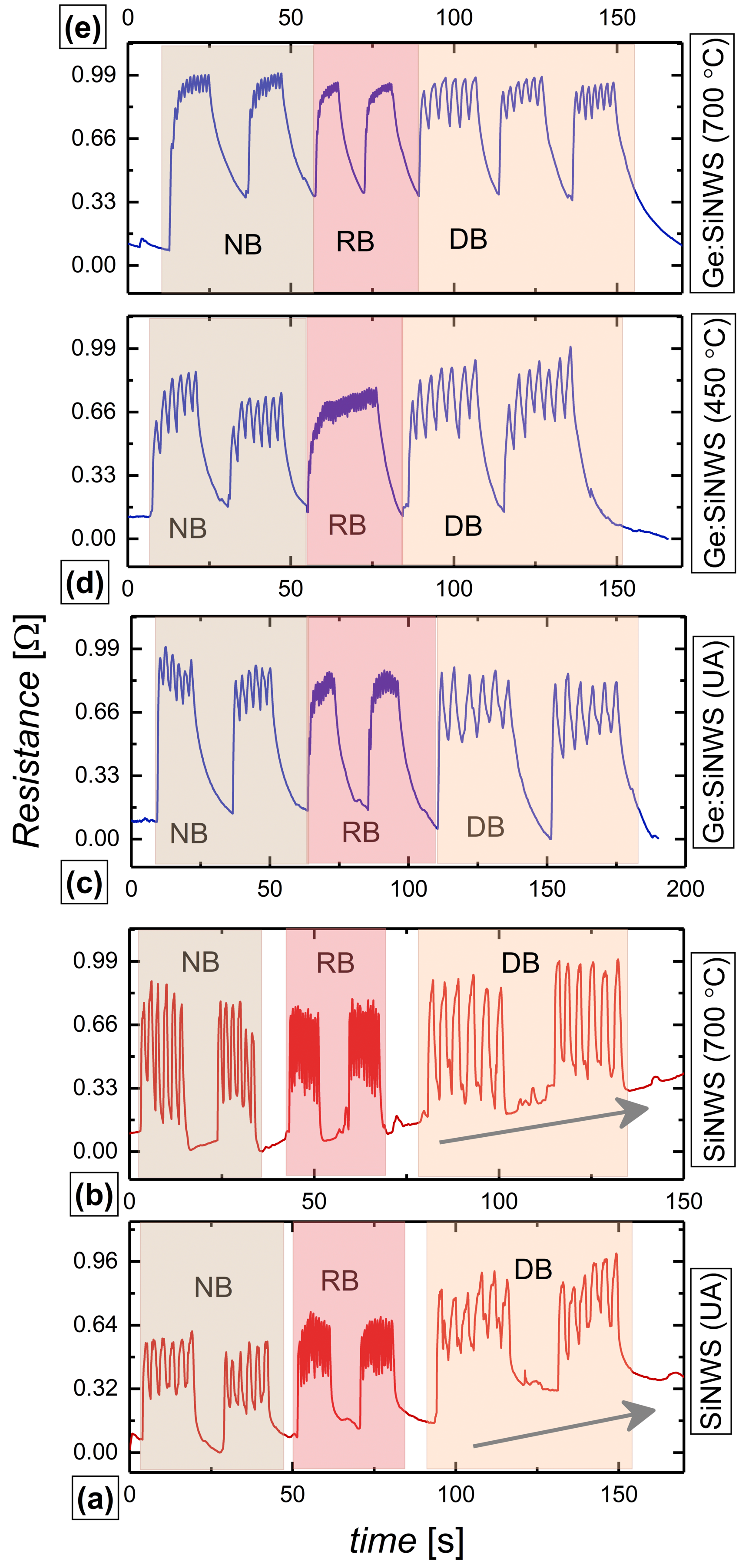}
    \caption{Room temperature resistance change as a function of time for (a) UA-SiNWs  (b) SiNWs annealed at 700$^{\circ}$C, (c) UA-Ge:SiNWs and annealed Ge:SiNWs at (d) 450 C and (e) 700$^{\circ}$C, respectively, under normal (NB), rapid(RB) and deep-breathing (DB) states, also represented by highlighted regions.}
    \label{3}
\end{figure}
\begin{figure}
    \centering
    \includegraphics[width=0.75\linewidth]{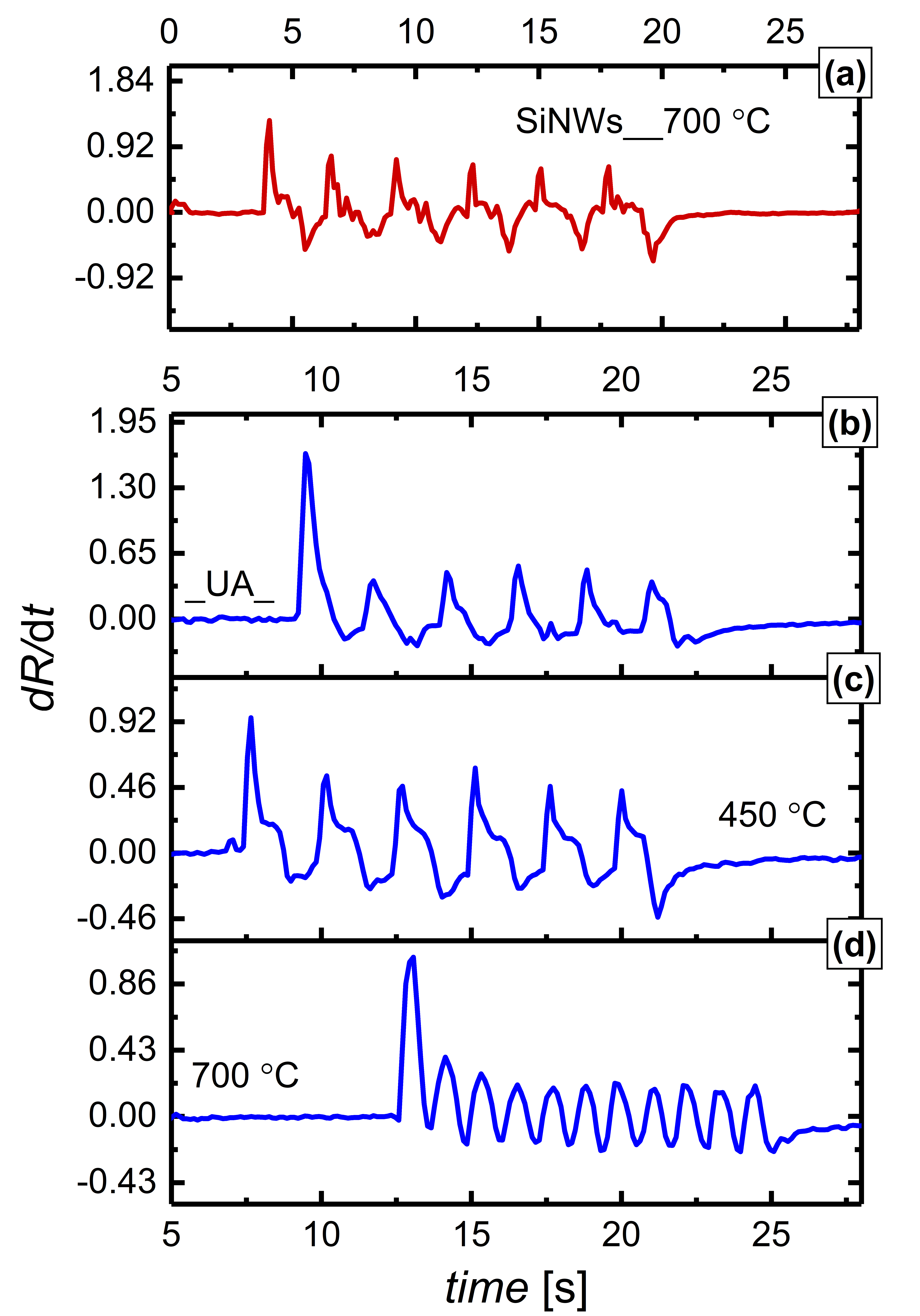}
    \caption{Differential plots obtained from Fig. \ref{3} of NB states for (a) SiNWs annealed at 700C$^{\circ}$C (b) UA-Ge:SiNWs and annealed Ge:SiNWs at (c) 450 $^{\circ}$C and (d) 700$^{\circ}$C, prospectively.}
    \label{4}
\end{figure}

\section{Experimental}
\subsection{Materials and methods}
Synthesis of random wire arrays (1cm~$^2$) of interconnected SiNWs were fabricated from 2$\times$2~mm$^2$ p-type, single-side polished 525~µm thick Si (001) substrates with resistivity of $\rho$ of 0.1-0.5~$\Omega$cm. Prior to synthesis the substrates were cleaned using acetone, methonal and isopropanol and DI water followed by nitrogen purging. Subsequently after drying the samples were etched with HF to remove any native oxides. Three step metal (silver, Ag) MACE process was utilized as follows:
\begin{enumerate}

\item The Si substrate were deposited with Ag nanoparticles catalyst by immersing the substrate in a solution of 3~M HF and 2 mM AgNO$_3$ for 60~seconds. 

\item To obtaine vertically aligned SiNWs, the substrates subsequently after deposition of Ag NPs, were etched by a HF:H$_{2}$O$_{2}$ (5M:0.4M) solution 

\item After etching the excess Ag nanoparticles were removed by immersing samples in a 20\% w/v HNO$_{3}$ solution.
\end{enumerate}

Two structural schemes were utilized in this study including SiNWs and Ge-coated SiNWs either in as-synthesized or annealed state. For the Ge deposition, direct current magnetron sputtering was utilized at constant power of 30 W using a 5N-Ge target. Argon was utilized as working gas and the throttle valves were adjusted to stabilize the growth pressure of 0.7 Pa. After deposition the sample were extracted for rapid thermal annealing and ex-situ characterizations. JPilec JetFirst 150 was utilized to anneal the samples at 450 and 700$^{\circ}$C under vacuum at $2\times 10^{-5}$ bar.
\subsection{Characterization}
The samples were characterized using X-ray diffraction (XRD) measurements using a Panalytical X’pert diffractometer (CuK$\alpha$, 0.15406 nm) and scanning electron microscopy (SEM,  Zeiss Supra 35). A detailed description regarding x-ray diffraction analysis is provided in our previous work \cite{Sultan2019}. For electrical measurements, two co-planar Au-contacts, 2×10 mm$^2$ each, with 250 nm thickness, were deposited on the surface of the samples via a hard mask using an electron beam evaporator (Polyteknik Cryofox Explorer 600 LT). The distance between the two contacts is 6 mm.

\section{Results and Discussion}
\subsection{Structure characterization}
Figure \ref{1}, shows (a) top view and (b) cross-sectional of SiNWs obtained after 20 min etching. The top-view image shows the wires are interconnected, forming a bundled rigid structure. Such bundle formation may take place because of capillary forces acting in the drying process following the wet-etching step.
The cross-sectional image shows that the
length of the wires is relatively homogeneous, around 5.5 $\mu$m.

The XRD diffractogram obtained over structures (i.e., SiNWs annealed and UA) around (004) atomic plane(Fig. \ref{2}), showed a variation in FWHM of peak, which is strongly influenced by strain-relaxation related  phenomena\cite{Romanitan2019}. For un-annealed (UA) SiNWs the XRD plot showed a peak along with broad hump, which when annealed at 700$^{\circ}$C showed a sharp feature. Such an effect can be attributed to structural defects and consequent strain relaxation phenomena in NWs which arise due to bending and torsion as has been well-documented in a study by Romanitan \textit{et. al.} \cite{Romanitan2019}. 

\subsection{Respiratory sensing}
The human respiratory sensing was investigated using various SiNWs based sensors (mentioned in experimental section), as shown in Fig. \ref{3}. All the structures were tested for three different breathing states i.e., normal, rapid and deep breathing and are labelled respectively in Fig. \ref{3}. It was observed that for SiNWs (both annealed and UA) there is a resistance baseline drift with time (indicated by arrow) unlike Ge-coated SiNWs in UA-state and those annealed at 450 and 700$^{\circ}$C showed significant reduction in baseline drift and can be attributed to higher Ge-hole mobility.

Additionally, although for bare SiNWs and those coated with Ge (UA and annealed at 450$^{\circ}$C) the change in resistance during breath waveforms is higher. However, the sensitivity to capture features of waveforms efficiently is poor, with lower response and recovery time, along with several unclassified non-periodic kinks during exhaling and inhaling breathing cycles. Such unwanted features in breathing profile can be visualized from the differential plots in Fig. \ref{4}(a-c). Furthermore, it was observed that compared to UA Ge-coated SiNWs, annealing the structure from 450 - 700$^{\circ}$C results in reduced oscillation intensity (i.e., change in resistance), which tends to be beneficial in gathering breathing information more efficiently at much faster response rate, as can be seen in Fig. \ref{4}(d), showing increased number of oscillation for a give time period of 14 s and with improved respiratory profile and absence of unwanted kinks or features.Considering the above scenario we have utilized Ge-coated SiNWs annealed at 700$^{\circ}$C for application in breath sensor.
\begin{figure}
    \centering
    \includegraphics[width=1\linewidth]{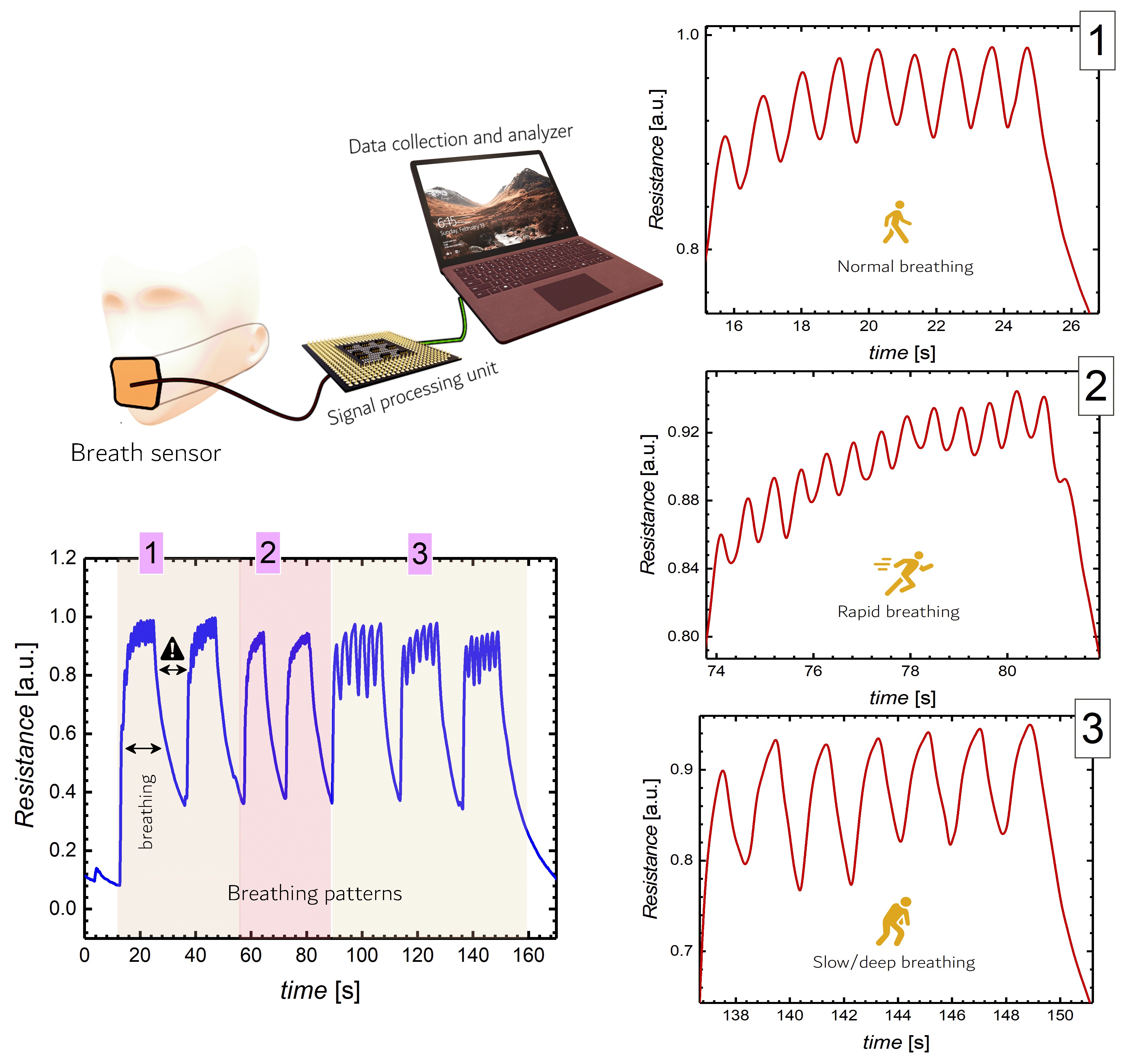}
    \caption{(a)Schemtic of human respiratory setup utilized. (b)The breathing response in real time recorded using annealed(700$^{\circ}$C) Ge:SiNWs. The insets marked as 1,2 and 3 corresponds different breathing states as highligthed in (b).}
    \label{5}
\end{figure}
The sample was mounted on a ceramic chip having patterned interdigital Au-electrode. Thin Al-wires were attached between co-planar contacts on sample and electrodes on ceramic chip. The sample was than mounted firmly on to philtrum using a double-sided tape, while the co-axial wires between the source meter and sample were properly dressed with in canula tube to avoid any loose connections and/or interference with unintended objects. A schematic illustration of setup is shown in Fig. \ref{5}(a).

\begin{figure}
    \centering
    \includegraphics[width=0.85\linewidth]{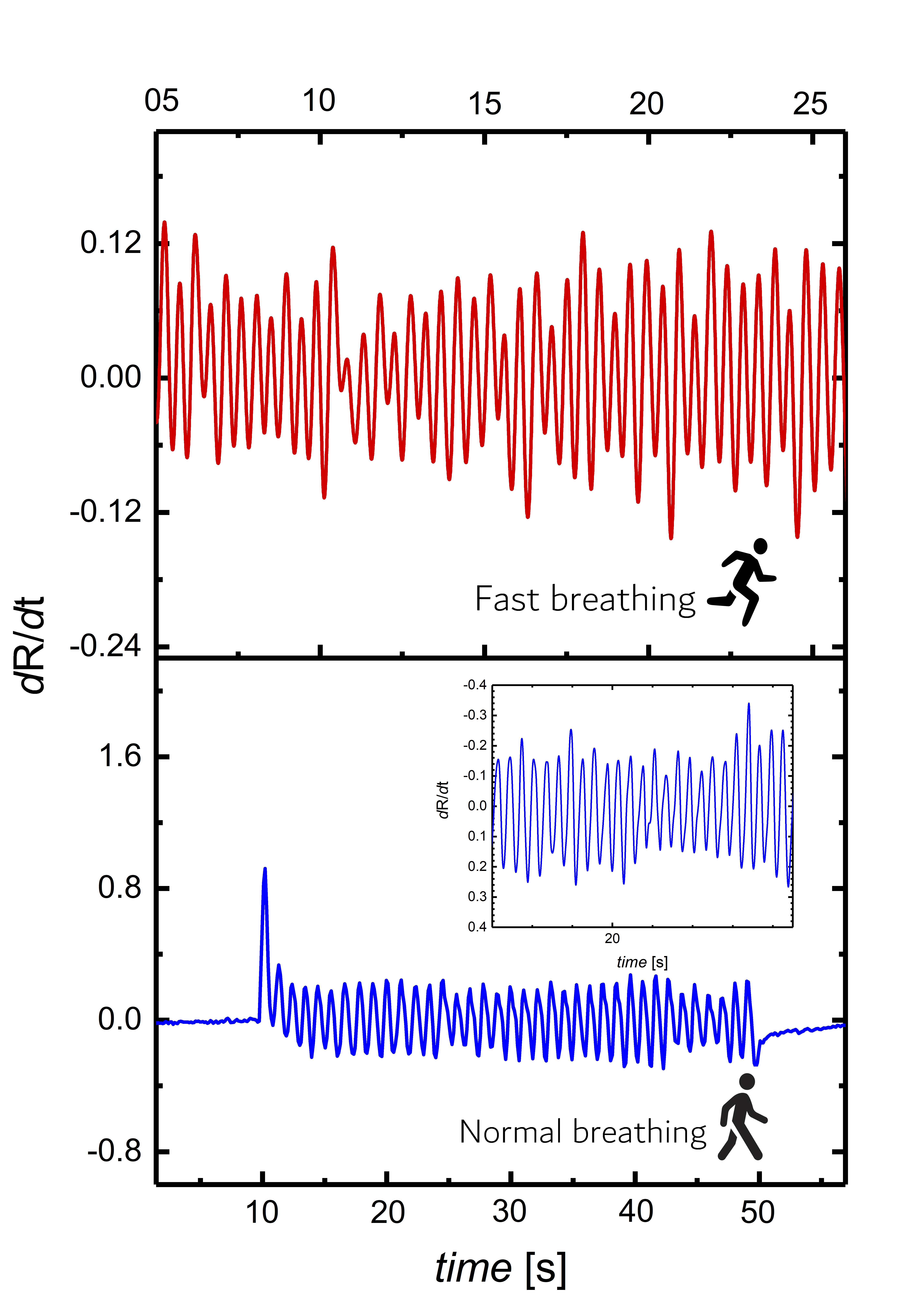}
    \caption{Differential plot for Ge:SiNWs structure annealed at 700$^{\circ}$C, 5 min, under NB(lower plot) and RB(upper plot) states for increased number of cycles. The sensor was tested again under similar condition for repeatability the results from which are shown in an inset for NB state.}
    \label{6}
\end{figure}

Fig. \ref{5}(b) shows the response of sensor (Ge coated SiNWs annealed at 700$^{\circ}$C) to periodic monitoring of the human breathing. It can be observed that the sensor can efficiently detect the different breathing states i.e., in our case normal, rapid and deep breathing (see insets in Fig. \ref{4}) 
compared to those with complex multi-step fabrication process and material systems \cite{lou2020human, Shi2021, He2018, Kundu2020} particularly composite fibres/polymers (such as PDMS) combined with conductive nano-materials like metals, CNTS, and graphene Which are often unstable and tend to degrade in ambient atmosphere over time. 
The sensor was able to capture breathing patterns without losing any features. It is to mention here that the waveform obtained during breathing is assumed to mimic the intended breath states while the a discontinuity between waveforms depicts a potential menace. Therefore, our sensors can provide a possible indication of risky situation for instance in case of sleep apnea or other breathing threats such as choking and asthma. Moreover, the structure was tested for re-usability by testing samples after storing it for a week and by testing sample for longer cycles and repeated breathing cycles. This is shown in Fig. \ref{6}, where a differential plot (dR/dt vs time) is shown anfor NB and RB sensing, where the inset in lower plot for NB shows the repeatability of sensor signal detection. It is to mention here that Fig. \ref{6} does not incorporate the entire breathing pattern recorded for sole purpose of figure readability.

\section{Conclusion}

In conclusion, we synthesized SiNWs by MACE for their application in respiratory sensing. The obtained structure were characterized via XRD and SEM showing SiNWs of $\sim$ 5.5 $\mu$m, bundled together. The structures were further treated either by annealing in RTA and/or deposited with  dc-sputtered Ge films with aim to increase the sensitivity and efficiency of sensor. It was observed that structure comprising of SiNWs deposited with Ge and annealed for 5 min at 700$^{\circ}$C resulted in higher efficiency, faster response and improved breathing profile, without any baseline drift in resistance. Further we have demonstrated the fabrication of portable, easy to fabricate and wearable senor with a great potential for application in devices to monitor the human respiratory profile and other possible application in healthcare facilities, moisture and pressure sensing.
 \\

\section*{Acknowledgment}

This work was supported by Reykjavik University Ph.D. fund no. 220006
and the Icelandic Research Fund Grant no. 218029-051.  
We are grateful to Rodica and Nucu Plugaru for discussions. \\

\bibliographystyle{ieeetr}
\bibliography{Bibliography}

\end{document}